\documentclass[useAMS,usenatbib]{mn2e}
\usepackage{graphicx}
\usepackage{txfonts}

\usepackage{natbib}
\title[Mass prediction using TCAF and POS models]{Estimation of Mass of the Black Hole Candidate
MAXI~J1659-152 using TCAF and POS Models}
\author[Molla et al.]
{Aslam Ali Molla$^1$, Dipak Debnath$^1$\thanks{E-mail: dipak@csp.res.in}, Sandip K. Chakrabarti$^{1,2}$, Santanu Mondal$^1$, Arghajit Jana$^1$ \\
$^1$ Indian Centre For Space Physics, 43 Chalantika, Garia Station Road, Kolkata, 700084, India\\
$^2$ S. N. Bose National Center for Basic Sciences, JD-Block, Salt Lake, Kolkata, 700098, India}

\begin{document}

\date{Accepted 2016 April 12, Received 2016 April 12; in original form 2015 December 17}

\maketitle

\begin{abstract}

The Galactic transient black hole candidate (BHC) MAXI~J1659-152 exhibited temporal and 
spectral evolution during its very first X-ray outburst (2010) after its discovery on 25th Sept. 2010. 
Our recent studies of a few transient BHCs including MAXI~J1659-152 using Chakrabarti-Titarchuk 
two-component advective flow (TCAF) solution as an additive table local model in XSPEC
revealed details of accretion flow dynamics around the black hole (BH). The TCAF model 
fitted normalization (N) comes out to be almost constant throughout the entire outburst 
consisting of several spectral states. We introduce two independent methods to determine the 
mass ($M_{BH}$) of the BHC, namely, {\it $i)$ keeping TCAF fitted normalization parameter in a 
narrow range}, and {\it $ii)$ studying evolution of the Quasi-Periodic Oscillation frequency 
($\nu_{QPO}$) with time, fitted with the propagating oscillatory shock (POS) model}. The predicted
mass ranges of the source with these two methods are $4.7-7.8~M_\odot$, and $5.1-7.4~M_\odot$ 
respectively. Combining results of these two methods, we obtain a most probable 
mass range of the source to be $4.7-7.8~M_\odot$ or $6^{+1.8}_{-1.3}~M_\odot$. 

\end{abstract}


\section{Introduction}

Transient compact sources, such as black holes (BHs) and neutron stars (NSs), have been studied extensively since the advent 
of X-ray astronomy, specially after the launch of {\it Rossi X-ray Timing Explorer (RXTE)} about two decades ago. Most of 
these systems are in binaries, comprising at least of one compact object, such as a NS or a BH 
as the primary which accretes matter which may be winds from their companions of matter overflowing the Roche-lobe. 
The spectral and temporal behaviors of BH and NS sources are very 
distinct in nature. Black hole candidates are uniquely identified by their masses, which are in excess of 
the Tolman–Oppenheimer–Volkoff (TOV) limit. Most of the observed black hole X-ray binaries (BHXBs) are transient 
in nature. These transient black hole candidates (BHCs) are primarily observed 
during their outbursts characterized by increased X-ray luminosity and successive transitions from 
one spectral state to another in a few days. In general, it has been found that these objects show
hard (HS), hard-intermediate (HIMS), soft-intermediate (SIMS) and soft (SS) spectral states
(see, Belloni et al., 2005; McClintock \& Remillard 2006, hereafter MR06 for a review). They also show
low- and high frequency quasi-periodic oscillations (QPOs) in power-density spectra (PDS) 
(see, Remillard \& McClintock, 2006, for a review) during their outbursts. These observed spectral 
states are also found to be strongly correlated with timing properties in hardness-intensity 
diagram (HID; see, Belloni et al. 2005; Debnath et al. 2008) or, in a more physical hysteresis diagram using
accretion rate ratio and X-ray intensity (ARRID; see, Mondal et al., 2014a; Jana et al., 2016). Different branches of 
these diagrams are found to be directly related to different observed spectral states, generally, in the sequence: 
HS $\rightarrow$ HIMS $\rightarrow$ SIMS $\rightarrow$ SS $\rightarrow$ SIMS $\rightarrow$ HIMS $\rightarrow$ HS. 

A large number of theoretical or phenomenological models exist in the literature, which claim to 
fit spectral data sets from these sources (MR06).  
With the inclusion of two-component advective flow (TCAF) model (Chakrabarti \& Titarchuk, 1995, 
hereafter CT95; Chakrabarti, 1997) in HEASARC's spectral analysis software package XSPEC 
(Arnaud, 1996) as a local additive table model (Debnath, Chakrabarti \& Mondal, 2014, hereafter DCM14; 
Mondal, Debnath \& Chakrabarti, 2014a, hereafter MDC14; Debnath, Mondal \& Chakrabarti, 2015a, hereafter DMC15; 
Debnath, Molla, Chakrabarti \& Mondal, 2015b, hereafter Paper-I; Jana et al., 2016, Chatterjee et al., 2016), we obtain a clear picture 
of accretion flow dynamics in several transient black hole candidates (e.g., H~1743-322, GX~339-4, MAXI~J1659-152, MAXI~J1836-194, MAXI~J1543-564) 
during their X-ray outbursts where the evolution of physical parameters, such as, the mass accretion rates of the Keplerian 
and sub-Keplerian flows, location and size of the Comptonizing cloud etc. are also obtained apart from usual flux 
and hardness ratio variations.

The Galactic transient BHXB MAXI~J1659-152 was discovered at the sky location of R.A. $= 16^h59^m10^s$, 
Dec $= -15^\circ 16'05''$ on 25th Sept. 2010 simultaneously by MAXI/GSC (Negoro et al., 2010) and SWIFT/BAT 
instruments (Mangano et al., 2010). The source showed its first main outburst phase for the duration of one and 
a half months only, other than a low-level activity (termed as the quiescence phase) which continued for around nine 
months after the main outburst phase. The source was extensively studied in multi-wave bands to explore both 
timing and spectral properties during the outburst and the quiescence phases. Kuulkers et al. (2010, 2013) 
predicted orbital periods of the binary system as $\sim 2.42$ hrs, which is the lowest so far among all BHXBs. 
They also predicted the companion of the object as an M5 dwarf star, consisting of mass $0.15-0.25~M_\odot$ and 
radius of $0.2-0.25~R_\odot$. So far, mass ($M_{BH}$) of the Galactic transient BHC MAXI~J1659-152 has not been measured 
dynamically. In the literature, one can find estimates ranging from $2.2-3.1~M_\odot$ (Kennea et al., 2011), 
to $3.6-8.0~M_\odot$ (Yamaoka et al., 2012), to $20\pm3~M_\odot$ (Shaposhnikov et al., 2011). 

Similar to other transient BHCs, different spectral states are observed during the entire phase of the outburst 
of this source. Low-frequency quasi-periodic oscillations (QPOs) and their evolutions are observed during 
the declining (HIMS and HS) phase in the same way. In Paper-I, we presented TCAF model fitted spectral results 
based on our detailed timing and spectral studies using RXTE proportional counter unit 2 (PCU2) data of 
Galactic transient BHC MAXI~J1659-152. We find three spectral states (HS, HIMS, SIMS) to be present during 
the entire phase of the 2010 outburst of MAXI~J1659-152 in the sequence of HIMS $\rightarrow$ SIMS 
$\rightarrow$ HIMS $\rightarrow$ HS. Soft state (SS) is missing during the outburst, which may be because 
of the lack of enough accretion rate of the Keplerian component to cool the hot Comptonizing region  
(so called `CENBOL' in TCAF solution). The whole work was carried out keeping 
the mass of the black hole frozen at $6M_\odot$ where a variable model $N$ in the range of $2.22-766.6$ was found.

In general, spectral fitted model $N$ (e.g., disk black body with power-law) vary from one observation 
to another. There are some reports of constant $N$ in data of some specific spectral states. For example, 
Steiner et al. (2010), fitting spectra of LMC~X-1 in soft states with a constant $N$. It was possible
because in soft states the inner edge of the Keplerian disk remains at around the marginally stable radius. 
In a robust model,  it should be left untouched since there is no scope for free parameters other than those 
arising out of governing equations (i.e., integral constants or paameters derived from them). This is because 
the $N$ is a function of mass, constant inclination angle and the distance, unless the disk precesses and the 
projected surface has a variable effective emission area along the line of sight. In that spirit, we investigate 
the dynamics with a near constant N, independent of the spectral states during the entire phase of the outburst 
with the hope to have a better estimate of the mass of the black hole itself while fitting the spectra with 
TCAF model. Since mass itself has error among other things, this normalization constant $N$ also has an error margin. 
Another goal would be to use the same normalization even for the next outbursts of the same source
and thus constraining the system parameters further. 
An interesting cross-check would be to plot the predicted mass range ($M_{BH}$) from this method as a function 
of fitted reduced $\chi^2$ ($\chi^2_{red}$). Best fitted $\chi^2_{red}$ vary with the fitting parameter mass of 
the BH ($M_{BH}$) while spectral fit with TCAF, and it was found to deviate from best fitted values ($\sim 1$). 
The minimum of the reduced $\chi^2$ of the $M_{BH}$-$\chi^2_{red}$ plot also provides us with a good estimation 
of the mass itself, if we limit ourselves in the range of $\chi^2_{red}$ as $\leq 1.5$.

One more independent method would be to study evolution of QPO frequencies during the rising and declining 
phases of the outburst. Chakrabarti and his collaborators have introduced a propagating oscillatory shock (POS) 
model (Chakrabarti et al., 2005, 2008; Debnath et al., 2010, 2013; Nandi et al., 2012), where $M_{BH}$ plays an 
important role in the governing equations. Thus, one can also predict the most probable mass range from the POS 
model fitted QPO frequency ($\nu_{QPO}$) vs. time (day) evolution (see, Iyer et al., 2015). In this paper, 
we predict the mass range of MAXI~J1659-152 from these two independent ways, one from spectral property and the 
other from the timing property. Interestingly, we find the ranges are almost identical.

This {\it paper} is organized in the following way: in the next Section, we define observation and data analysis 
procedure. In \S 3, a summary of the results of our analysis using TCAF and POS model fits are discussed. We
show how two approaches could be used to determine mass ranges of the BHC MAXI~J1659-152. Finally, 
in \S 4, we discuss the limitations and relative credibility of our approach to predict mass of an unknown 
black hole and make concluding remarks.

\section{Observation and Data analysis}

RXTE has covered the entire 2010 outburst of MAXI~J1659-152 starting from 2010 September 28 (Modified Julian Day, i.e., MJD=55467) 
to 2010 November 10 (MJD=55508) roughly on a daily basis. We analyze archival data of the RXTE PCA instrument using HEASARC's 
software package HeaSoft version HEADAS 6.15 and follow the standard data analysis techniques (Debnath et al. 2013, 2015a,b) 
to analyze the PCA data. We have used the PCA Event mode data with a maximum timing resolution of 125$\mu$s for timing analysis. 
In order to generate power-density spectra (PDS), we apply the command ``powspec'' of XRONOS package with a normalized factor 
of '-2' to get the expected 'white' noise subtracted rms function variability on $2.0-15$~keV (0-35 channels of PCU2) light curves 
of $0.01$~sec time bins. Unit of power is rms$^2$/Hz. Lorentzian profiles are used to fit power density spectra (PDS) to 
find centroid frequency of QPOs and "fit err" command is used to find errors in QPOs.

For the spectral analysis, we follow the same techniques as discussed in Paper-I. We fit the background subtracted spectra with 
TCAF based model {\it fits} file within an energy range of 2.5-25 keV. During the entire 
outburst we use $3.0 \times 10^{21}~atoms~cm^{-2}$ as the value of hydrogen column density (N$_{H}$) 
as proposed in Mu\~{n}oz-Darias et al. (2011) for the absorption model 
{\it wabs}. We assume a fixed 1\% systematic instrumental error for the spectral study during the entire phase of the outburst.
To obtain the BH spectra with the current Version (v0.3) of the TCAF, one needs to supply a total of five input 
parameters. The parameters are : $i)$ mass of black hole ($M_{BH}$) 
in solar mass ($M_\odot$) unit, $ii)$ sub-Keplerian rate ($\dot{m_h}$ in $\dot{M}_{Edd}$) unit, 
$iii)$ Keplerian rate ($\dot{m_d}$ in Eddington rate $\dot{M}_{Edd}$) unit, $iv)$ location of the shock which is really the
inner edge of the Keplerian component and the outer edge of the Compton cloud or CENBOL  
($X_s$ in Schwarzschild radius $r_s$=$2GM_{BH}/c^2$) unit, $v)$ compression ratio ($R$) of the shock, which is 
the ratio of post-shock density to the pre-shock density ($\rho_+/\rho_-$). The model normalization ($N$) is a 
fraction $\frac{r_s^2}{4\pi D^2} sin(i)$, where `$D$' is the source  distance  (in units of $10$~kpc) and 
`$i$' is the disk inclination angle. In Paper-I, we kept the mass of the BH frozen at $6~M_\odot$ and allowed 
to vary $N$. Here, however, we keep all the input parameters of the TCAF model free while analyzing 
all the $30$ observations in order to obtain mass itself.

\section{Models and Results}

We consider two different approaches to estimate the mass of MAXI J1659-152. 
These methods are discussed in the following sub-sections: (\S 3.1) {\it Constant Normalization parameter approach}, 
and (\S 3.2) {\it QPO frequency ($\nu_{QPO}$)-Time (day) evolution fitted with POS model}.
The estimated mass ranges from these two different methods are also discussed. Finally,
we combine these results to have a reasonable estimation of the mass of the source.

\subsection{Constant TCAF Model Normalization Method}

In Paper-I, during the spectral fitting of MAXI J1659-152 with TCAF model by keeping mass of black hole frozen at 
6$M_\odot$, we found that the value of normalization constant varied over a wide range of $2.22-766.6$. This was done 
following conventional models where the variation of normalization could be attributed to variation 
of disk thickness at the inner disk just outside the Compton cloud or CENBOL. However, physically it is preferable 
to have a constant normalization since the TCAF solution has no scope of any free parameter other
than the five inputs. Normalization depends on mass, distance and angle, which are constant in all observations, 
but they are also unknown. It also depends on the instrument response function and the absorption by the
intervening medium. It is also to be noted that TCAF normalization depends on $M_{BH}$ is a non-linear way. 
Because of these complexities, we allow the fits to vary the mass and the normalization ensuring that both 
remain within narrow ranges as allowed by satisfactory $\chi^2$ fit. 
If the normalization is found to change abruptly in some days it would mean that there are additional
components, such as jets whose contributions are not included in the present version (v0.3) of the 
TCAF model {\it fits} file. We find that the model normalization remains roughly 
constant in the range of $129.7-146.3$ for reasonable fits throughout the outburst. 
Variations of extracted flow parameters such as $\dot{m_h}$, $\dot{m_d}$, $X_s$, $R$ show roughly similar nature as 
seen in Paper-I. Most interestingly, we see that model fitted mass of the BH comes within 
a range of $4.2 M_\odot$ - $7.7 M_\odot$. Insensitivity of flow temperatures 
on the mass of the black hole is the main reason for this large range. 
In Table 1, model fitted normalization values and mass values for all the $30$ observations are presented. 
We also fitted these observational data by freezing the normalization constant at $139.07$, an average of
normalization (see, Col. 3 of Table 1) obtained when used above as a free parameter. 
Here, we get the mass of the BH within a narrower range of $4.7 M_\odot$ - $7.8 M_\odot$. 
In Fig. 1(a-c), variations of TCAF model fitted shock location (in $r_s$), normalization and mass values with time (Day in MJD) 
are shown, when all model parameters were assumed to be free. Figure 1d shows variation of model fitted 
masses when the normalization was frozen at an average value of $139.07$.

\begin{figure}
\vskip 0.8cm
        \centerline{
	\includegraphics[scale=0.6,width=8truecm,angle=0]{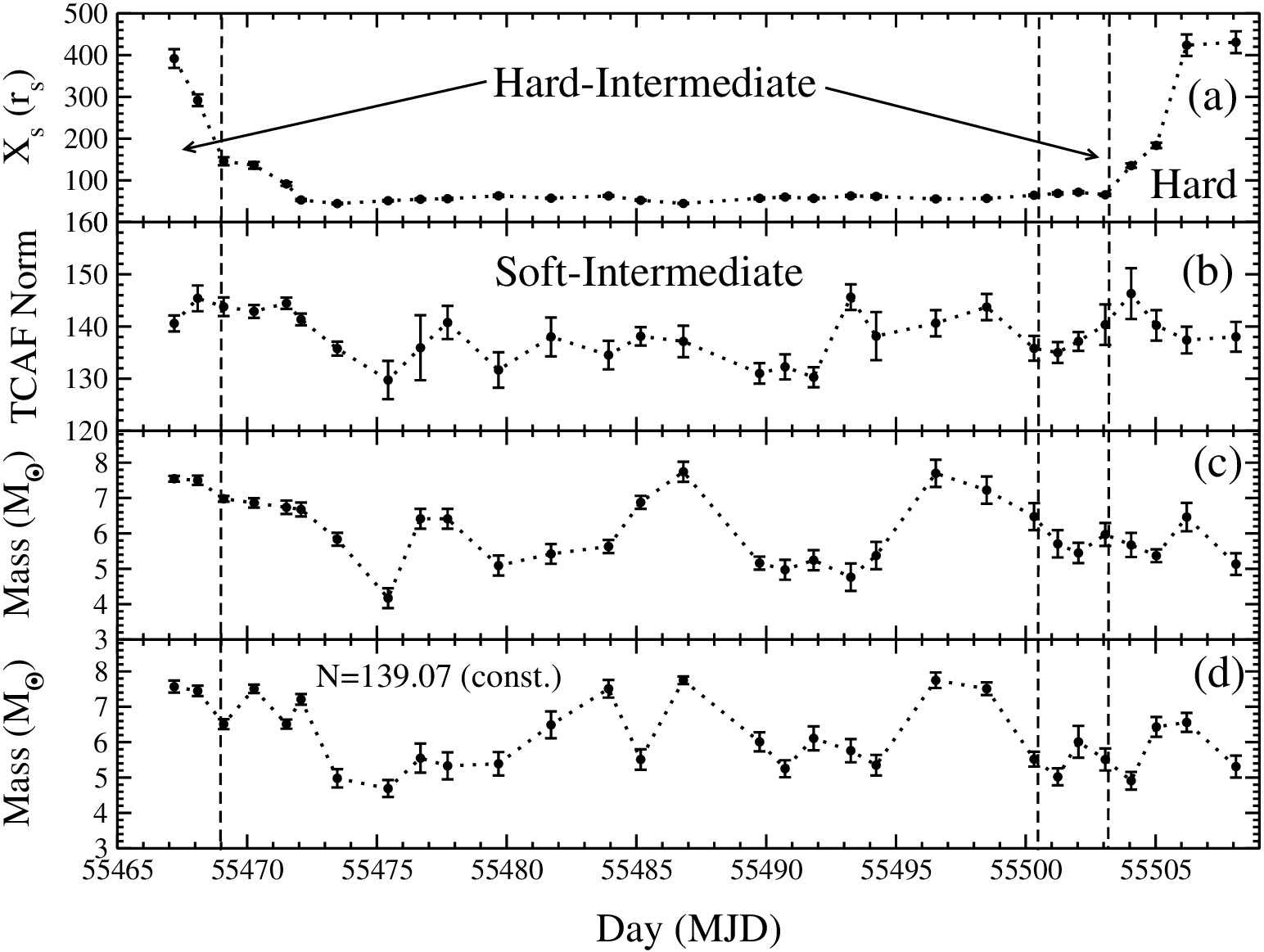}
        }
\caption{Variations of TCAF model fitted parameters: (a) shock location (in $r_s$), (b) normalization (N), and 
(c) mass of the black hole ($M_{BH}$ in $M_\odot$) over the entire period of the 2010 outburst of MAXI~J1659-152 
are shown when all model parameters are kept as free.  
In (d), TCAF fitted $M_{BH}$ is shown when model normalization was kept frozen at $139.07$, 
which we get from taking an average of the normalization values of panel (a).}
       \label{fig1}
\end{figure}

We can also cross-check the range of the predicted mass by the above method when
we vary $M_{BH}$ vs. reduced $\chi^2$ of the best fitted spectra. 
The best fit using the current version (v0.3) of TCAF model was obtained based on 
the model fitted reduced $\chi^2$ values when found nearly equal to unity. In Paper-I and in Table 1, 
model parameters are given when best fits are obtained. After getting satisfactory 
model fits, we kept all model parameters, such as sub-Keplerian rate $\dot{m_h}$, Keplerian rate $\dot{m_d}$, 
location of the shock $X_s$, compression ratio $R$, normalization $N$ as frozen, except mass of the 
black hole and found how reduced $\chi^2$ changed with mass of black hole. In Fig. 3(b-d), 
variation of reduced $\chi^2$ with mass of BH for three different spectra (obs. Ids: 95118-01-06-01, 
95118-01-16-01, 95118-01-21-00) selected from three different spectral states (SIMS, HIMS and HS respectively) 
are shown. As in the earlier cases, if we consider $\chi^2_{red}=1.5$ to be the upper limit for acceptable fits, 
we obtain the mass to be in the range of $4.4 M_\odot - 7.3 M_\odot$. 

\begin{table}
	\centering
	\caption{TCAF Model Fitted Mass and Normalization Values}
	\label{tab:table1}
	\begin{tabular}{lccccc} 
	\hline
   Obs. & Id      &   Day     &  Norm. (N)  &  Mass         &  Mass     \\ 
        &         &  (MJD)    &             & $(M_\odot)$   &  $(M_\odot)$ \\ 
    (1) &   (2)   &    (3)    &   (4)       &  (5)          &  (6)  \\
\hline                                                                              
 1& X-02-00  &  55467.19 & $140.6^{\pm 1.53}$  & $7.55^{\pm 0.08}$    & $7.57^{\pm 0.17}$ \\
 2& X-02-01  &  55468.09 & $145.3^{\pm 2.47}$  & $7.11^{\pm 0.13}$    & $7.45^{\pm 0.15}$ \\
 3& X-02-02  &  55469.09 & $143.8^{\pm 1.82}$  & $6.98^{\pm 0.08}$    & $6.51^{\pm 0.14}$ \\
 4& X-03-00  &  55470.26 & $142.9^{\pm 1.32}$  & $6.87^{\pm 0.13}$    & $7.51^{\pm 0.12}$ \\
 5& Y-03-00  &  55471.51 & $144.5^{\pm 1.11}$  & $6.74^{\pm 0.19}$    & $6.51^{\pm 0.13}$ \\
 6& Y-05-00  &  55472.07 & $141.3^{\pm 1.22}$  & $6.68^{\pm 0.19}$    & $7.21^{\pm 0.15}$ \\
 7& Y-09-00  &  55473.47 & $135.8^{\pm 1.35}$  & $5.84^{\pm 0.18}$    & $4.98^{\pm 0.26}$ \\
 8& Y-13-00  &  55475.43 & $129.7^{\pm 3.67}$  & $4.17^{\pm 0.28}$    & $4.69^{\pm 0.24}$ \\
 9& Y-17-00  &  55476.67 & $135.9^{\pm 6.22}$  & $6.42^{\pm 0.28}$    & $5.55^{\pm 0.41}$ \\
10& Y-19-00  &  55477.72 & $140.8^{\pm 3.18}$  & $6.41^{\pm 0.28}$    & $5.33^{\pm 0.38}$ \\
11& Y-23-00  &  55479.68 & $131.7^{\pm 3.38}$  & $5.09^{\pm 0.28}$    & $5.39^{\pm 0.33}$ \\
12& Y-27-00  &  55481.71 & $138.1^{\pm 3.72}$  & $5.42^{\pm 0.28}$    & $6.49^{\pm 0.38}$ \\
13& Y-30-00  &  55483.92 & $134.5^{\pm 2.74}$  & $5.63^{\pm 0.18}$    & $7.51^{\pm 0.25}$ \\
14& Z-02-00  &  55485.16 & $138.1^{\pm 1.77}$  & $6.88^{\pm 0.18}$    & $5.51^{\pm 0.29}$ \\
15& Z-03-00  &  55486.80 & $137.1^{\pm 3.04}$  & $7.74^{\pm 0.28}$    & $7.75^{\pm 0.11}$ \\
16& Z-06-01  &  55489.74 & $131.1^{\pm 1.95}$  & $5.16^{\pm 0.18}$    & $6.01^{\pm 0.27}$ \\
17& Z-07-00  &  55490.72 & $132.3^{\pm 2.41}$  & $4.97^{\pm 0.28}$    & $5.25^{\pm 0.24}$ \\
18& Z-09-00  &  55491.82 & $130.3^{\pm 1.92}$  & $5.24^{\pm 0.28}$    & $6.11^{\pm 0.34}$ \\
19& Z-10-00  &  55493.25 & $145.6^{\pm 2.46}$  & $4.76^{\pm 0.39}$    & $5.76^{\pm 0.33}$ \\
20& Z-11-00  &  55494.23 & $138.2^{\pm 4.61}$  & $5.38^{\pm 0.38}$    & $5.35^{\pm 0.29}$ \\
21& Z-13-00  &  55496.53 & $140.6^{\pm 2.51}$  & $7.71^{\pm 0.38}$    & $7.75^{\pm 0.22}$ \\
22& Z-15-00  &  55498.49 & $143.7^{\pm 2.51}$  & $7.22^{\pm 0.38}$    & $7.51^{\pm 0.18}$ \\
23& Z-16-00  &  55500.31 & $135.8^{\pm 2.36}$  & $6.48^{\pm 0.38}$    & $5.52^{\pm 0.21}$ \\
24& Z-16-01  &  55501.23 & $135.1^{\pm 2.01}$  & $5.71^{\pm 0.39}$    & $5.02^{\pm 0.24}$ \\
25& Z-17-00  &  55502.02 & $137.2^{\pm 1.81}$  & $5.45^{\pm 0.29}$    & $6.01^{\pm 0.45}$ \\
26& Z-17-01  &  55503.06 & $140.4^{\pm 3.89}$  & $5.97^{\pm 0.33}$    & $5.51^{\pm 0.31}$ \\
27& Z-18-00  &  55504.06 & $146.3^{\pm 4.86}$  & $5.67^{\pm 0.34}$    & $4.91^{\pm 0.25}$ \\
28& Z-19-00  &  55505.03 & $140.2^{\pm 2.93}$  & $5.37^{\pm 0.18}$    & $6.43^{\pm 0.28}$ \\
29& Z-20-00  &  55506.20 & $137.4^{\pm 2.55}$  & $6.46^{\pm 0.40}$    & $6.56^{\pm 0.27}$ \\
30& Z-21-00  &  55508.09 & $138.1^{\pm 2.86}$  & $5.13^{\pm 0.31}$    & $5.31^{\pm 0.31}$ \\

\hline
\end{tabular}
\noindent{
\leftline {Here X=95358-01, Y=95108-01, and Z=95118-01 are the initial part of } 
\leftline {the observation Ids. In Cols. 4 and 5, TCAF fitted model normalization and }
\leftline {BH mass values are shown when all TCAF related parameters are free } 
\leftline {while fitting the spectra, and in Col. 6, model fitted mass values are shown}
\leftline {when normalization is frozen at $139.07$.} 
\leftline {Note: Here, $1~\sigma$ errors for N and masses are shown in superscripts.}
}
\end{table}

\begin{figure}
\vskip 0.8cm
        \centerline{
	\includegraphics[scale=0.6,width=8truecm,angle=0]{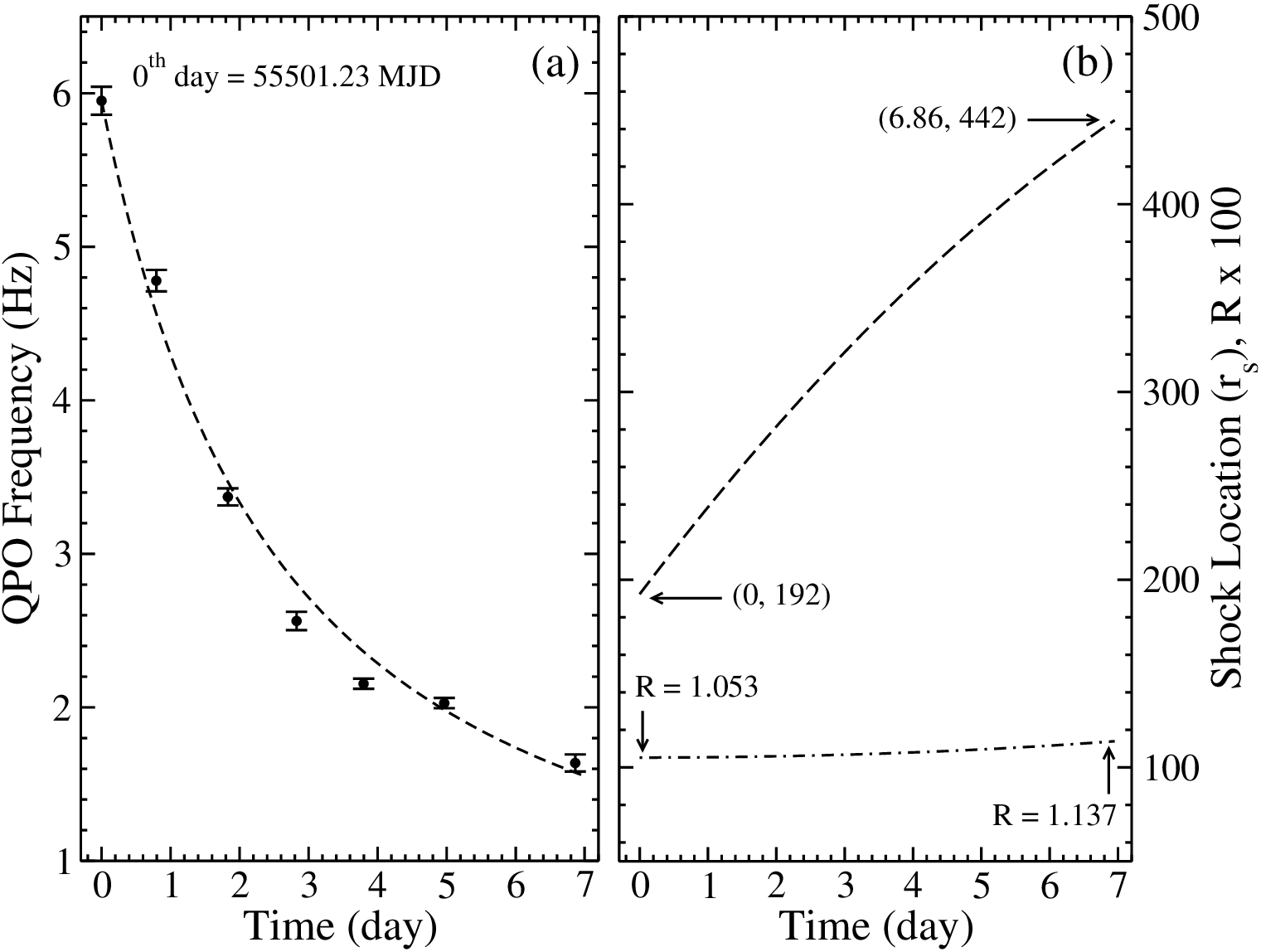}
        }
\caption{(a) Variation of QPO frequency with time (in day) during the declining phase of the 2010 
outburst of MAXI~J1659-152, fitted with the POS model solution (dashed curve). In plot (b), variation 
of POS model fitted shock locations (in $X_s$) and compression ratios are shown.}
       \label{fig2}
\end{figure}

\subsection{\it{QPO Frequency($\nu_{QPO}$)-Time(day) Evolution: Fitted with POS Model}}

Low frequency QPOs are very commonly seen in power density spectra of X-ray intensity variations (light curves) 
emitted from disks surrounding stellar massive black holes. Sometimes dominating frequency of these QPOs (generally,
type `C') are found to evolve with time, mostly in HS and HIMS from both rising and declining phases 
of an outburst. The monotonic evolution of these QPO frequencies could be fitted very well 
by a propagating oscillatory shock (POS) model (see, Chakrabarti et al., 2005, 2008, 2009; Debnath et al., 2010, 2013; 
Nandi et al., 2012). According to POS, the frequency is inversely proportional to the in fall time scale from the 
shock location $X_s$. In fitting with TCAF, $X_s$ is considered to be an input parameter. The main governing equations 
of POS model are (Chakrabarti et al., 2008; Debnath et al., 2013): Equation for in fall time scale ($t_{infall}$):
$$
t_{infall}\sim  X_s/v \sim  R X_s(X_s-1)^{1/2},
\eqno{(1)}
$$
where, $X_s$ is the shock location in units of the Schwarzschild radius $r_s=2GM_{BH}/c^2$, $v$ is the velocity of 
propagating shock wave in $cm/sec$. $R$ is the compression ratio. Both $X_s$ and $R$ are free parameters for TCAF fits.

Equation for instantaneous QPO frequency ($\nu_{QPO}$ in $sec^{-1}$) is:
$$
\nu_{QPO} = \nu_{s0}/t_{infall}= \nu_{s0}/[R X_s (X_s-1)^{1/2}], 
\eqno{(2)}
$$
where, $\nu_{s0}= c/r_s=c^3/2GM_{BH}$ is the inverse of the light crossing time of the BH of mass $M_{BH}$ 
in unit of $sec^{-1}$ and $c$ is the velocity of light. 

Equation for instantaneous shock location ($X_s(t)$):
$$
X_s(t)=X_{s0} \pm v_0 t/r_s ,
\eqno{(3)}
$$
where, $X_{s0}$ is the shock location at time $t = 0$ (the day when the evolving QPO is observed for the first time) and
$v_0$ is the corresponding shock velocity in the laboratory frame. The positive sign in the second term
is to be used for an outgoing shock in the declining phase and the negative sign is to be used for the
in-falling shock in the rising phase. 

Depending on the rate of cooling in the post-shock region which drives the variation of shock location, 
shock may be accelerating or decelerating: $v(t)=v_0 \pm a t$, where $v_0$ is the initial velocity 
and $a$ is the acceleration of the shock front. Accordingly, the shock compression ratio $R$ 
(= $\rho_+$/$\rho_-$, where $\rho_+$ and $\rho_-$ are the densities in the post- 
and the pre- shock flows) also may vary in the following way:
$1/R \rightarrow 1/R_0 \pm \alpha (t_d)^2$, where $R_0$ is the compression ratio of the first day, 
$t_d$ is the time in days (assuming first observation day as 0$^{th}$ day) and $\alpha$ is a constant which 
determines how the shock (strength) becomes weaker/stronger with time. $\alpha$ is positive when shock becomes 
weaker with time (rising phase) and negative when the shock gets stronger (declining phase). 

This POS model has been successfully applied to study evolution of QPO frequencies during rising and 
declining phases of the outbursts of a few transient BHCs, such as GRO~J1655-40 (Chakrabarti et al., 2005, 2008), 
XTE~J1550-564 (Chakrabarti et al., 2009), GX~339-4 (Debnath et al., 2010; Nandi et al., 2012), 
H~1743-322 (Debnath et al., 2013) and IGR~J17091-3624 (Iyer et al., 2015). Iyer et al. (2015) also 
showed that the mass of an unknown BHC can be predicted from the POS model fitted QPO frequency evolutions, 
since in POS the mass is an important parameter deciding the shock distance $X_s$ (see, Eq. 2). 
We also apply this to study observed type `C' QPO frequency evolution of MAXI~J1659-152 during its declining 
phase of the 2010 outburst with POS model. The evolution (monotonically increasing) of type `C' QPOs 
(from $1.61$ to $2.72$~Hz) during initial rising phase (HIMS) of the outburst are observed only for 
three days (from MJD=55467.19 to 55469.09), which is indeed hard to use for determination of mass. 
During SIMS, as in other transient BHCs, type B or A QPOs are observed sporadically whose 
origin may be different from the resonance condition (see, Chakrabarti et al., 2015 for more details).

During the declining phase of the 2010 outburst of MAXI~J1659-152, QPO frequency is found to decrease 
monotonically from $5.95$~Hz (Obs. Id: 95118-01-16-01, and MJD=55501.23) to $1.63$~Hz (Obs. Id: 95118-01-21-00, 
and MJD=55508.09) in about $\sim 6.86$~days. From the best fitted POS model (see, Fig. 2), we observe that 
during the evolution, the shock recedes with a deceleration of $\sim 35$~m/sec/day. The shock velocity decreases 
from $1000$~cm/sec to $759$~cm/sec during this period and the shock moved away from the BH from $\sim 192 r_s$ to 
$\sim 442~r_s$. In the same time, $R$ is found to be increase slightly, starting from $\sim 1.05$ to $\sim 1.14$ 
with a constant $\alpha = -0.0015$. This type of slow movement (in few $m/sec$) of the shock wave agrees quite 
well with many observational results reported earlier (Chakrabarti et al., 2005, 2008, 2009; 
Debnath et al., 2010, 2013; Nandi et al., 2012) as well as theoretical results 
(Mondal et al., 2015) for other transient BHCs during their QPO evolutions. It is to 
be noted that for the best fitted POS model, the mass of the BHC was found 
to be at $6~M_\odot$. This was a frozen parameter in Paper-I to fit BH spectra with TCAF model solution. 
So, it appears that our choice of constant mass value of $6~M_\odot$ in Paper-I to fit BH spectra with TCAF 
was sufficiently good. The values of all POS model fitted parameters along with calculated and observed QPO 
frequencies are given in Table 2. 

We have changed $M_{BH}$ in POS model equation and tried to refit QPO frequency evolution with the modified 
POS solutions and found that model fitted values of the reduced $\chi^2$ deviate from its best fitted value 
at $6~M_\odot$. The variation of model fitted reduced $\chi^2$ values with $M_{BH}$ is shown in Fig. 3a. 
Now, if we restrict ourself to the reduced $\chi^2$ value = 1.5 for the best fit, the boundary 
of the mass of the BHC should be in the range of $5.1-7.4~M_\odot$.

\begin{figure}
\vskip 0.8cm
\centerline{
\includegraphics[scale=0.6,width=8truecm,angle=0]{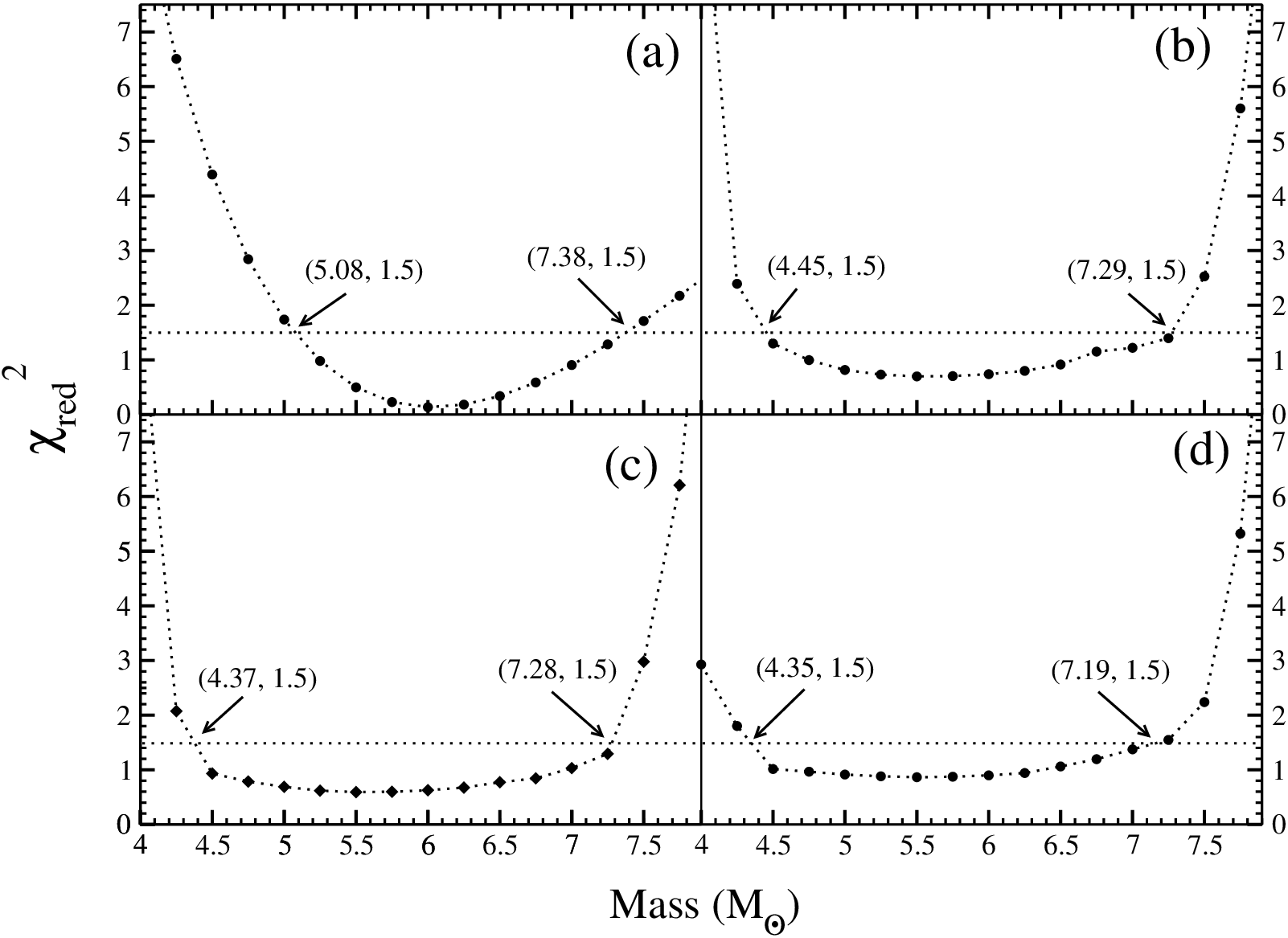}
}
\caption{ (a) Variation of the mass using the POS fitted QPOs. Reduced $\chi^2$ remains within acceptable limit 
($\leq 1.5$), for the predicted mass range of $5.1-7.4$~$M_\odot$. (b-d) Variation of TCAF model fitted reduced 
$\chi^2$ ($\chi^2_{red}$) with mass of the BH ($M_{BH}$ in $M_\odot$) for three different observations 
(Obs Ids: 95118-01-06-01, 95118-01-16-01, 95118-01-21-00), selected from three different spectral states, 
such as SIMS, HIMS, and HS respectively. Here, we obtain preferable mass range of $4.4-7.3$~$M_\odot$. 
for acceptable reduced $\chi^2$ limit.}
\label{fig3}
\end{figure}

\begin{table}
\centering
\caption{Declining Phase QPO evolution : Fitted with POS Model}
\label{tab:table1}
\begin{tabular}{lcccccccc} 
\hline
 Obs. &	Time &      $\nu$   &      $\nu_{th}$ & $X_s$   & v       & R\\
      &  (day)&      (Hz)    &       (Hz)      & $(r_s)$ &(cm/s)   &   \\ 
\hline
24& 0.000 &     5.951 &       5.953 &      192.3 &   1000.1 & 1.053\\
25& 0.793 &     4.779 &       4.565 &      229.3 &   972.2 &  1.054\\
26& 1.829 &     3.371 &       3.469 &      274.5 &   935.9 &  1.058\\
27& 2.826 &     2.563 &       2.807 &      314.5 &   901.1 &  1.066\\
28& 3.796 &     2.154 &       2.363 &      350.3 &   867.1 &  1.077\\
29& 4.965 &     2.028 &       1.984 &      389.2 &   826.2 &  1.095\\
30& 6.863 &     1.638 &       1.576 &      442.6 &   759.8 &  1.137\\
\hline
\end{tabular}
\noindent{
\leftline{$\nu$ is the observed QPO frequency, $\nu_{th}$ is the theoretical QPO}
\leftline{frequency calculated from POS model fit, $X_s$ is the shock location in} 
\leftline{Schwarzschild radius ($r_s$), `$v$' is the velocity of shock in $cm/sec$,} 
\leftline{and `$R$' is the shock compression ratio.} 
}
\end{table}

\section{Discussion and Concluding Remarks}
In this paper, our aim has been to determine the mass of the Galactic transient BHC MAXI J1659-152 using two aspects of TCAF paradigm, namely, 
from spectral and temporal studies. 
We use the data of RXTE PCA instrument during its very first outburst. 
So far, the mass of this source was not measured dynamically, although in the literature one can find reports of
the predicted mass in the range of $2.2 - 8.0 ~M_\odot$ (Kennea et al., 2011; Yamaoka et al., 2012). 
We use two independent methods such as {\it $i)$ 
carrying out spectral fit using data of the entire range of the outburst, and 
\it $ii)$ QPO frequency ($\nu_{QPO}$)-time (day) evolution, fitted with POS model}.
These methods lead to a reasonably  narrow range of the mass of the Galactic transient BHC MAXI J1659-152, which 
is more close to the range predicted by Yamaoka et al. (2012). Shaposhnikov et al. (2011) obtained very 
high mass ($\sim  20~M_\odot$) probably because the applicability of the limiting QPO frequency method may be 
questionable as we do not see any soft state in this object.

Ideally, since the mass, the distance and the inclination angle are all supposed to be constants, the 
normalization factor on the overall spectrum should also remain a constant when the data is 
already corrected for the instrumental response and the absorption due to interstellar medium. However,
none of these quantities are known accurately and RXTE resolution ($\sim 20$\%) is not high enough to get 
accurate spectra. Thus in our fits, we vary both the mass and the normalization 
factor and find that they remain within narrow ranges for reasonably good reduced $\chi^2$. 
Uncertainty in the mass is mainly due to insensitivity of temperature of the disk and Compton cloud 
on mass. Other factors are the error bars of the data and the consequent allowance in reduced $\chi^2$ values. 
In the present context, by leaving the mass as a free parameter, we find that the normalization ($N$) 
remains in a very narrow range of $129.7$ to $146.3$ and mass of the BH comes in the range of 
$4.2-7.7~M_\odot$. If we take an average of $N$ mentioned above, we obtain  $N \sim 139.07$. 
Freezing this number as TCAF model normalization, we refit for all $30$ observations, and find 
that mass range of source comes out to be $4.7-7.8~M_\odot$ (see, Col. 6 of Table 1). 
Interestingly, the variations of TCAF model fitted/derived physical flow parameters 
remain similar to what were reported in Paper-I. Ratio of halo accretion rate and disk accretion rate (ARR) attains 
a local maximum on exactly the same day when HIMS $\rightarrow$ HS transition is observed during 
declining phase of the outburst. In models such as disk blackbody (diskbb) and power-law  
normalization constants in both the components are allowed to vary arbitrarily and thus, in a sense 
they are also free parameters. In our fit with TCAF solution, we can use only five free parameters including the mass,
and hence  the freedom is quite restricted. Even then, we find the mass, variation of the accretion rates
along with the variation of the Compton cloud size to be quite reasonable and as a whole we obtain a very 
good picture of accretion flow behaviour during an outburst. Such a knowledge would help us understanding 
how the viscosity must have changed at the outer edge in order to generate such a variation of the flow parameters.

If we proceed to the other mass determination method (QPO frequency evolution study using POS model), 
the predicted mass range is found to be in agreement with mass range 
mentioned above. In the second method, we studied evolution of the dominating QPO frequency during declining phase 
of the outburst with POS model (Chakrabarti et al., 2005, 2008, 2009; Debnath et al., 2010, 2013; Nandi et al., 2012; 
Iyer et al., 2015). We get the predicted mass range to be $5.1-7.4~M_\odot$ from the QPO evolution study.

It was instructive to see how reduced $\chi^2$ varies with the mass of the black hole. So we plotted this variation and obtain
a range after restricting acceptable reduced $\chi^2$ to be below $1.5$. In Fig. 3a,
the most probable mass range comes out to be $5.1-7.4~M_\odot$ 
with a minimum of $M_{BH}$ at $\sim 6~M_\odot$. We also see a similar minimum
nearly at $M_{BH}=6~M_\odot$ for three different spectra selected from three different spectral states 
(Fig. 3b-d). These plots give us a range of $4.4-7.3~M_\odot$. So, this gives us a 
consistency check. 

In summary, we can conclude that after combining the results from these methods, the predicted mass range is
to be $4.7-7.8~M_\odot$ for Galactic transient BHC MAXI~J1659-152.
Since in POS model fit of QPO frequency evolution during declining phase of the outburst, we got the best fitted 
$\chi^2_{red}$ for $M_{BH}=6~M_\odot$, we could further conclude that the mass of the BH to be
$6^{+1.8}_{-1.3}~M_\odot$. 

Finally, we should mention that in the current TCAF model {\it fits} file (v0.3), we have not included effects 
due to spin of Kerr BHs. We believe that spin affects the features very close to the horizon, i.e., in the softest 
states when the inner edge of the Keplerian disk is close to the marginally stable orbit. Otherwise, the shock 
locations are smaller for the same initial flow parameters. Inclusion of spin would reduce the derived mass also 
since some spin energy remains distributed in space time to cause dragging of inertial frames. But we did not have 
any softest state here and we obtained the mass using properties away ($X_s > 44~r_s$) from the black hole. So for 
this particular BHC MAXI~J1659-152, the result we derived may not change even when spin is included. In near future, 
we will include this important spin parameter and its effects in the modified version of the model {\it fits} file. 

\section*{Acknowledgements}
AAM and SM acknowledge supports from MoES sponsored junior and post-doctoral research fellowships respectively. 
DD acknowledges support from project fund of DST sponsored Fast-track Young Scientist (SR/FTP/PS-188/2012). 
AJ and DD acknowledges support from ISRO sponsored RESPOND project (ISRO/RES/2/388/2014-15) fund.


{}


\end{document}